\documentclass[a4paper,fleqn,usenatbib,referee,epsfig,color]{mnras}

\usepackage[T1]{fontenc}
\usepackage{ae,aecompl}

\usepackage{graphicx}	% Including figure files
\usepackage{amsmath}	% Advanced maths commands
\usepackage{amssymb}	% Extra maths symbols

\newif\ifAMStwofonts
\newcommand{\be}{\begin{equation}}
\newcommand{\ee}{\end{equation}}
\newcommand{\bea}{\begin{eqnarray}}
\newcommand{\eea}{\end{eqnarray}}

\begin{document}
\title[Halo Magnetic field Structure]{\bf  Exact spirally symmetric galactic dynamos }
\author[R.N. Henriksen, A. Woodfinden, J.A. Irwin]
{R. N. Henriksen$^1$\thanks{henriksn@astro.queensu.ca},
A. Woodfinden$^1$\thanks{17aw14@queensu.ca}, and J. A. Irwin$^1$\thanks{irwinja@queensu.ca}\\
%Q. D. Wang$^4$\thanks{wqd@astro.umass.edu}\\
$^1$Dept. of Physics, Engineering Physics \& Astronomy, Queen's University, Kingston, Ontario, K7L 3N6, Canada\\}
\date{Accepted XXX. Received YYY; in original form ZZZ}
\pubyear{2017}
% Don't change these lines
%\begin{document}
\label{firstpage}
\pagerange{\pageref{firstpage}--\pageref{lastpage}}
\maketitle

\begin{abstract}
This paper extends the results of an earlier paper on scale invariant  galactic spiral magnetic modes to  time dependent, scale invariant, spirals. The examples given are all exact in that they may be described in terms of hypergeometric functions. We restrict the discussion to an infinitely conducting medium in order to avoid  earlier approximation, which limited the solutions to cones lying within about twenty degrees to the plane. The magnetic disc spirals, `X type" poloidal fields, and the recent discovery of rotation measure screens in edge-on galactic halos were all recovered in such solutions. 
\end{abstract}
\newpage

\section{Introduction}

In this article we present  analytic solutions for spirally symmetric magnetic spiral arms based on the assumption of scale invariance {\it in classical dynamo theory}. In \cite{Hen2017b} modal solutions for the magnetic field were developed in a steady state although, but for one analytic example, the results were restricted to lie on cones of large vertex angle (surfaces close to the disc). Under this approximation the classical dynamo equations were reduced to homogeneous linear equations for the spiral modes. The condition for a non-trivial solution provided a relation between the `dynamo number', mode number, scale invariant velocity field and similarity `class'. This allowed for qualitative comparison with observations of  rotation measure (RM) such as those found in \cite{CMP2016}, and with face-on and edge-on magnetic fields (\cite{Beck2015},\cite{Kr2015} .  

The assumption of the steady state avoids quenching and growth considerations, but  restricts the kind of solutions available since generation of flux must be strictly balanced by resistive diffusion and advection. Only one  exact solution was offered in \cite{ Hen2017b} and this was not extensively developed. In this article we turn to a {\it time dependent, scale invariant, approach}. Such a formulation allows slightly more flexibility in the analytic solutions, while the assumption of temporal scale invariance  automatically gives a growing or decaying magnetic field. This temporal evolution may eventually be compared to present field strengths and galactic lifetimes, in order to constrain initial galactic field strengths. The solution is self-similar in time so that the spatial geometry of the field is independent of epoch. The field strength is not independent of epoch, but it depends only on time dependent multiplicative constants.

This paper is in the same spirit as a companion paper that treats only axial symmetry \cite{HWI2018} both as a steady state and  with time dependence. In particular, although the data  are not yet in an available form that can be  fitted by the methods of this paper, we believe that our examples display common, if not generic, behaviour in either axial or non axial symmetry.

Although the solutions found in this paper are analytic, they are exact only in that they may be identified as hypergeometric functions. These yield essentially a  series in the arc tangent of the angle with the plane, which  is similar in spirit to the approach in \cite{Hen2017b} and \cite{Hen2017}. Nevertheless, under quite general conditions the various series converge to yield exact solutions and so the magnetic halo may be explored to arbitrarily  large angle with the disc. 

{\it Under  the assumption of zero resistive diffusion (infinite conductivity)}, we  reduce the complete scale invariant problem to two equations for the azimuthal components of the vector potential and the magnetic field. Together these two equations show that the restricted problem is second order in the azimuthal vector potential. These equations are readily solvable numerically for the vector potential in combination. If  resistive diffusion is included, the equations can only be reduced to a fourth order equation in the azimuthal vector potential. We record this general case for future reference, but work only with the simpler case. 

The scale invariant symmetry in all of these papers is taken to extend to the velocity field, and to the sub-scale  resistivity and helicity. This avoids many poorly understood physical assignments, but it also by-passes halo dynamics at any spatial scale.  The ultimate physical justification lies in any agreement with observed features, as will be noted briefly below. However, complex interacting systems do often become scale invariant in regions remote from boundaries, including the initial state (\cite{Barenblatt96},\cite{Hen2015}). This would restrict the application of our halo fields  to regions well away from the  galactic bulge, and from any dominant interaction with the intergalactic medium. For example, although halo lag might also be scale invariant (e.g. \cite{HI2016})  and so compatible with our model, interactions with other galaxies would not be. 

Although restricted physically by the scale invariant assumption, we find that observed magnetic behaviour as reported in \cite{Beck2015}, \cite{Kr2015} and \cite{CMP2016} is present in these examples. Moreover they produce magnetic field geometries that have been empirically developed in \cite{FT2014} and inferred for the Milky Way in \cite{Gfarr2015}. Our approach is independent of much previous  detailed simulation and theoretical analysis  that is based on various physical mechanisms. Many of these agree with the  magnetic geometry deduced here. Some relevant references are \cite{KF2015}, \cite{Black2015}, \cite{B2014}, \cite{M2015}, \cite{MS2008} and \cite{SS1990}. A wealth of information is becoming available due to the CHANG-ES project mapping the halos of edge-on galaxies (e.g. \cite{WI2015}. This edge-on data, together with the face-on magnetic spirals (\cite{Beck2015}) and ultimately even the observations of the Milky Way \cite{Hv2015},\cite{Val2017} will either confirm or infirm the scale invariance.

%\section{Exact Scale Invariant Spiral Dynamos}

\section{Steady Dynamo  Magnetic Fields with Zero Resistive Diffusion?}

One may ask whether scale invariant, steady,  solutions exist without resistive diffusion. This restriction to an infinitely conducting medium allows  analytic, non axially symmetric, solutions to be found (below) with time dependence. 
However  we have found as a general conclusion that {\it no steady, scale invariant, solutions exist}, at least in the absence of global electric fields. The demonstration follows by writing the steady dynamo equation (equation (\ref{eq:Afield}) with $\partial_t=0$) with $\eta=0$. The terms that remain are a sum of the convective term, which is perpendicular to the magnetic field, and of the sub-scale dynamo term, which is parallel to the magnetic field.  Clearly there is no non-trivial solution for the magnetic field that makes the sum equal to zero.

More formally, if we write the steady  equation (\ref{eq:Afield}) in terms of the scaled magnetic and velocity fields when $\eta=0$ (see definitions below), we find that there are no derivatives of the magnetic field in the resulting three homogeneous equations. Setting the determinant of these equations equal to zero in order to have a non trivial solution, gives (see the definitions below)
\be
\bar\alpha_d^2+\bar{\bf v}^2=0,
\ee
which has no real solution. The conclusion is the same  either with or without axial symmetry.

%This conclusion would not be true were the condition 
%\be
%\alpha_d^2+{\bf v}^2=0,
%\ee
%to be satisfied. Here $\alpha_d$ is the usual sub-scale helicity and ${\bf v}$ is the plasma velocity, as used again below. This relation could  hold if, say, $\alpha_d$ were to be complex. In such a case the  dynamo equations with zero resistivity would have to be solved separately for the real and imaginary parts of the vector potential, which are no longer identical.  Ultimately these equations would yield the amplitude and phase  of the magnetic field at each  spatial point.  It is of reasonable physical interest to introduce phase shifts between the mean magnetic field and the sub scale generation, both steady and time dependent, but we will not pursue this here.

\section{Scale Invariant, time dependent, spirally symmetric dynamos}

We refer to the classical mean-field dynamo equations  \cite{M1978} in the form for the  magnetic vector potential \cite{Hen2017b}
\be
\partial_t{\bf A}={\bf v}\wedge \nabla\wedge {\bf A}-\eta\nabla\wedge\nabla\wedge {\bf A}+\alpha_d\nabla\wedge {\bf A}.\label{eq:Afield}
\ee
In these equations ${\bf v}$ is the mean velocity, $\eta$ is the resistive diffusivity and $\alpha_d$ is the magnetic `helicity' resulting from sub-scale magnetohydrodynamic turbulence and ${\bf A}$ is the potential. We proceed by setting $\eta=0$ so that the medium is infinitely conducting. 

Under the assumption of temporal scale invariance, the time dependence will simply be a power law or (in the limit of zero similarity class) an exponential factor. Hence the geometry of the magnetic field remains `self-similar' over the time evolution, and we can therefore study the geometry without requiring a fixed epoch.  Although we have  no way of bringing the dynamo into a  finite steady state, the compatible time evolution of $\alpha_d$, ( and $\eta$ when  that is retained)  and ${\bf v}$ is  also given by the scale invariance. 

It is important in what follows to observe that the time derivative in equation (\ref{eq:Afield}) is taken at a fixed spatial point. We do not therefore differentiate the unit vectors. This study differs from that of the axially symmetric case by including  a dependence of the magnetic field on azimuthal angle. This dependence, plus the possible rotation  in time of the magnetic field (e.g.  moving magnetic spiral arms), is included through the  scale invariant variable $\Phi$  to be defined below. The azimuthal dependence is restricted to be of the spiral form by using the combined variable $\kappa$, which is also defined below.   

The scale invariance is found following the technique advocated in \cite{CH1991} and \cite{Hen2015}.  We first introduce a time variable $T$ along the scale invariant direction according to 
\be
e^{\alpha T}=1+\tilde\alpha_d\alpha t,\label{eq:T}
\ee
where $\tilde \alpha_d$ is a numerical constant that appears in the scale invariant form for the helicity, $\alpha_d$, which form is to be given below.  The constant  numerical factor $\tilde\alpha_d$  in equation (\ref{eq:T}) is purely for subsequent notational convenience. The quantity $\alpha$ should not be confused with the helicity as it is an arbitrary scale used in the  temporal scaling. The cylindrical coordinates $\{r,\phi,z\}$ are transformed into scale invariant variables $\{R,\Phi,Z\}$ according to (e.g. \cite{Hen2015})\footnote{We take spatial variables to be measured in terms of a fiducial Unit such as the  radius of the galactic disc.} 
\be
r=Re^{\delta T},~~~~\Phi=\phi+(\epsilon +q)\delta T,~~~~z=Ze^{\delta T},\label{eq:scaled variables}
\ee
where $\delta$ is another arbitrary scale that appears in the spatial scaling, and $\epsilon$ is a number that fixes the rate of rotation of the magnetic field in time. We add $q$ to the arbitrary $\epsilon$ for subsequent algebraic convenience (see equation (\ref{eq:kappa}) below). In our subsequent discussion $1/q>0$ appears as the pitch angle of a  spiral mode that is lagging relative to the sense of increasing angle $\phi$. \footnote{It should be noted that in \cite{Hen2017b}, $q$ had this r\^ole as the {\it normally} defined pitch angle with respect to the azimuth. In our examples $tan^{-1}(1/q)$ is typically $tan^{-1}(0.4)\approx 22^\circ$.} We note that 
\be
e^{\delta T}=(1+\tilde\alpha_d\alpha t)^{1/a},\label{eq:delta}
\ee
where the `similarity class' $a\equiv \alpha/\delta$ is a parameter of the model defined as the self-similar `class' \cite{CH1991}, which reflects the Dimensions of a global constant. This is discussed in some detail in \cite{HWI2018}, but a simple example is afforded by a global constant $GM$ where $G$ is Newton's constant and $M$ is some fixed mass. This is the situation for Keplerian orbits. The space-time Dimensions of $GM$ are $L^3/T^2$ and after scaling length by $e^{\delta T}$ and time by $e^{\alpha T}$, $GM$ scales as $e^{(3\delta-2\alpha)T}$. To hold this invariant under the scaling we must set $\alpha/\delta\equiv a=3/2$, which is the `Keplerian  class'.  Note that this `class' , that is the ratio $3/2$ of the powers of spatial scaling to temporal scaling gives Kepler's third law, $L^3\propto T^2$.

 The constant $\tilde \alpha_d$ determines the strength of the sub-scale dynamo.  
%such as the reciprocal rotation period of the galaxy. 

As is usual in this series of presentations we write the magnetic field as 
\be
{\bf b}=\frac{\bf B}{\sqrt{4\pi \rho}},\label{eq:b}
\ee
so that it has the Dimensions of velocity. Here $\rho$ is not associated with the dynamo and indeed might have the value $1/(4\pi)$ in cgs Units, but it is completely arbitrary. It may in fact  be absorbed into the  multiplicative constants that appear in our solutions. 

In temporal scale invariance the variable fields must have the form
\bea
{\bf A}&=&\bar{\bf A}(R,\Phi,Z)e^{(2-a)\delta T},\nonumber\\
{\bf b}&=& \bar{\bf b}(R,\Phi,Z)e^{(1-a)\delta T}\equiv \nabla\wedge {\bf A},\nonumber\\
{\bf v}&=& \bar{\bf v}(R,\Phi,Z)e^{(1-a)\delta T},\label{eq:tempfields}
 \eea   
where the barred quantities are the scale invariant fields. 
 %The quantity $a\equiv \alpha/\delta$ and denotes the `Similarity class' as defined in \cite{CH1991}. This is essentially the ratio of the spatial to temporal powers that appear in the Dimensions of some globally conserved quantity. Thus for example, Keplerian orbits about some mass $M$ preserve $GM$ where $[GM]=L^3/T^2$ so that $a=3/2$. We do not have to define the global constant in advance using our procedure. 

Considering equations (\ref{eq:tempfields}) and equation (\ref{eq:delta}) we see that the time dependence is generally a power law in powers of $(1+\tilde\alpha_d \alpha t)$ , where the power is determined by the `class' parameter $a$. The time scale is set by the value of $1/(\tilde\alpha_d\alpha)$. Should $\alpha=0$ (which implies a global constant with the Dimension of time) we find from equation (\ref{eq:delta}) that  $\delta T=\tilde\alpha_d \delta t$. the field can then  grow exponentially according to equations (\ref{eq:tempfields}). The helicity, velocity field and indeed  the diffusivity will grow correspondingly. The time scale is controlled by the value of $1/(\tilde\alpha_d\delta)$, which may be long.

The helicity arising from the sub-scale $\alpha_d$, and the resistive diffusivity $\eta$, must be written  according to their respective Dimensions as 
 \bea
 \alpha_d&=&\bar\alpha_d(R,\Phi,Z)e^{(1-a)\delta T},\nonumber\\
\eta&=& \bar \eta(R,\Phi,Z) e^{(2-a)T}.\label{eq:tempparams}
\eea
We include the diffusivity here in order to indicate the necessary form of the scale invariance when it is present \footnote{The general equations with resisitve diffusion are straight forward to find, but they do not concern us here.}, but we set it equal to zero in the examples of this paper. 

At this stage a substitution of the forms (\ref{eq:tempfields}) into equations (\ref{eq:Afield}) yields three partial differential equations in the variables $\{R,\Phi,Z\}$. Thus, only the time dependence has been eliminated (e.g. \citet{CH1991}) through the assumption of temporal scale invariance. However we are seeking non axially symmetric spiral symmetry in the magnetic fields to match the observations summarized in \cite{Beck2015} and \cite{Kr2015}. Any combination of the  scale invariant quantities $\{R,\Phi,Z\}$ will  render the barred quantities in equations (\ref{eq:tempfields}) scale invariant, so we are free to seek a spiral symmetry by combining them. 

We choose a combination inspired by our previous modal analysis \cite{Hen2017b}. We assume that the angular  dependence may be combined with $R$  in a  {\it rotating} logarithmic spiral form as  (recalling the definition of $\Phi$ from equation (\ref{eq:scaled variables}))
\be
\kappa\equiv \Phi+q\ln{R}\equiv \phi+q\ln(r)+\epsilon T.\label{eq:kappa}
\ee
Moreover we combine the $R$ and $Z$ dependence into  a dependence on the conical angle through 
\be
\zeta\equiv \frac{Z}{R}.\label{eq:zeta}
\ee
The linearity of equations (\ref{eq:Afield}) allows us to seek solutions in the complex form 
\be
  {\bf \bar A}(R,\Phi,Z)={\bf \tilde A}(\zeta)e^{im\kappa}.\label{eq:fieldform}
\ee

On substituting these assumed forms into equation (\ref{eq:Afield}) one finds that a solution is possible in terms of $\kappa$ and $\zeta$,  {\it provided that  the ancillary quantities satisfy }
\bea
\bar\alpha_d&=& \tilde\alpha_d\delta R,\nonumber\\
\bar\eta&=& \tilde\eta \delta R^2,\label{eq:params}\\
{\bf \bar v}&=&\tilde \alpha_d \delta R~\{u,v,w\}.\nonumber
\eea 
The constant quantities  denoted $\tilde{()}$ and the velocity components $\{u,v,w\}$ are Dimensionless.

 Under these conditions the equations (\ref{eq:Afield}) become, {\it after setting the resistive diffusion equal to zero},
\bea
K\tilde A_r-im\tilde A_z&=& (1+imq)v\tilde A_\phi-(1+\zeta v)\tilde A'_\phi-w(\tilde A'_r+\zeta \tilde A'_z-imq\tilde A_z),\nonumber\\
(K-imv)\tilde A_\phi&=& -u(1+imq)\tilde A_\phi+(\zeta u-w)\tilde A'_\phi+im(w\tilde A_z+u\tilde A_r)+(\tilde A'_r+\zeta \tilde A'_z-imq\tilde A_z),\nonumber\\
im\tilde A_r+K\tilde A_z&=& (1+imq)\tilde A_\phi+(v-\zeta)\tilde A'_\phi+u( \tilde A'_r+\zeta \tilde A'_z-imq\tilde A_z),\label{eq:explicitA}
\eea
Where the prime indicates differentiation with respect to $\zeta$ and 
\be
K\equiv (2-a)+im(\epsilon+v).\label{eq:K}
\ee
The coherence of these equations (assuming physical solutions) in spirally symmetric,  scale invariant form,  {\it already predicts the existence of  rotating spiral magnetic dynamo fields}.

The magnetic field that follows from the curl of the potential takes the form
\be
\bar{\bf b}=\frac{\tilde{\bf b}}{R}e^{(im\kappa)},\label{eq:bbar}
\ee
where
\bea
{\bf\tilde b}&=&\{im\tilde A_z-\tilde A'_\phi,~~\tilde A'_r+\zeta \tilde A'_z-imq\tilde A_z,~~(1+imq)\tilde A_\phi-\zeta\tilde A'_\phi-im\tilde A_r\},\label{eq:btilde}\\\ &\equiv& \{\tilde b_r,\tilde b_\phi, \tilde b_z\}\nonumber 
\eea
Equations (\ref{eq:btilde}),(\ref{eq:bbar}), and the second of equations (\ref{eq:tempfields}) together {\it give the complete time dependent magnetic field}. 

%We note that  the form of $\tilde b_\phi$,  as expressed in terms of the components of the vector potential, appears in all three of equations (\ref{eq:explicitA}). One may therefore eliminate this combination of components  from equations (\ref{eq:explicitA}) by calling it $\tilde b_\phi$. This requires adding the vector potential expression for $\tilde b_\phi =\tilde A'_r+\zeta \tilde A'_z-imq\tilde A_z$  to the set (\ref{eq:explicitA}) as a fourth equation.   

An examination of equations (\ref{eq:explicitA})  indicates that one can rewrite equations (\ref{eq:explicitA}) as one second order equation for $\tilde A_\phi$. The algebra is however formidable. One effective procedure is to replace with $\tilde b_\phi$  the combination $\tilde A'_r+\zeta \tilde A'_z-imq \tilde A_z$  everywhere in equations (\ref{eq:explicitA}). We emphasize that the resulting equations do not `know'  that this combination of potentials is in fact the azimuthal field. We might have called the combination $X$.  

Subsequently we use the first and third equations of the set (\ref{eq:explicitA}) to solve for $\tilde A_r$ and $\tilde A_z$ in terms of $\tilde b_\phi$, $\tilde A_\phi$ and $\tilde A'_\phi$. Substituting these expressions for $\tilde A_r$ and $\tilde A_z$ into both the expression for $\tilde b_\phi$ following from the second equation of  equations (\ref{eq:explicitA}) and into the form of $\tilde b_\phi$ in terms of the potentials from equation (\ref{eq:btilde}),  yields after tedious algebra  two equations for $\tilde b_\phi$ and $\tilde A_\phi$ in the form
\bea
\tilde b_\phi(K^2-m^2(1+u^2+w^2))&=& (K-imv)\big[(K^2-m^2+(Ku-imw)(1+imq))\tilde A_\phi\nonumber\\
&+&((w-u\zeta)K+im(u+w\zeta))\tilde A'_\phi\big],\label{eq:bphi1}
\eea
\bea
 \tilde b_\phi\big(K^2&-&m^2(1+qw)+imqKu\big)+\tilde b'_\phi\big((K-im\zeta)w-(K\zeta+im)u\big)=\nonumber\\
&-&(K-imv)\big((1+\zeta^2)\tilde A''_\phi-2imq\zeta\tilde A'_\phi+imq(1+imq)\tilde A_\phi\big)\label{eq:bphi2}
\eea
One must exercise caution in using these two equations. Rather than treating them as two equations for the quantities $\tilde A_\phi$ and $\tilde b_\phi$ , the correct procedure is to substitute the first into the second in order to obtain a second order differential equation for $\tilde A_\phi$.  The resulting equation  is rather elaborate, given a general velocity field, so that it is more convenient to make the substitution after a particular velocity field has been chosen. Hence writing the two equations as above is in reality an algebraic aide, with the first equation  an intermediate step. 

%This peculiarity is due to the fact, as was remarked earlier, that the first equation is  the azimuthal equation (second of equations (\ref{eq:explicitA})) combined with the first and third equations once solved for $\tilde A_r$ and $\tilde A_z$ as functions of $\tilde b_\phi$, $\tilde A_\phi$ and $\tilde A'_\phi$. This procedure never really identifies $\tilde b_\phi$ in terms of the vector potentials, but the second equation does. 

Subsequently the potentials $\tilde A_r$ and $\tilde A_\phi$  can be found from the first and third equations of (\ref{eq:explicitA}) in the form 

\bea
(&K&-imuw)\tilde A_r-im(1+w^2)\tilde A_z=\nonumber\\
&[&(1+imq)(v-uw)-w(K-imv)]\tilde A_\phi-[1+v\zeta+w(w-u\zeta)]\tilde A'_\phi,\label{eq:AR}
\eea
and 
\bea
im(1+u^2)\tilde A_r+(K+imuw)\tilde A_z&=&[(1+imq)(1+u^2)+u(K-imv)]\tilde A_\phi\nonumber\\
&+&[v-\zeta+u(w-u\zeta)]\tilde A'_\phi.\label{eq:AZ}
\eea
Once again we leave the explicit linear solution for $\tilde A_r$ and $\tilde A_z$ for specific cases of the velocity field for simplicity in the presentation. Once these are found in terms of the solution of equation (\ref{eq:bphi2}) for $\tilde A_\phi$,
all of the field components (including the azimuthal component in terms of $\tilde A_r$ and $\tilde A_z$) follow from the expressions in equations (\ref{eq:btilde}) and (\ref{eq:bbar}).

In the next section we give a series of time dependent examples that are of interest in making comparisons with observations. One apparent simplification 
allows the vertical velocity to vary on cones according to $w=u\zeta$.  This does not change equations ((\ref{eq:AR}),(\ref{eq:AZ})), or the intermediate equation (\ref{eq:bphi1}), but the  equation (\ref{eq:bphi2}) for $\tilde A_\phi$ adds the term 
\be
(K-im\zeta)u,\label{eq:term}
\ee
to the bracket multiplying $\tilde b_\phi$.

\section{Exact Examples of Time Dependent Dynamo Spiral Magnetic Fields}

We look at some simple cases that illustrate generic properties. Specific fits to observational data will require more extensive parameter searches.

\subsection{Only Rotation in the Pattern Frame}

In \cite{Hen2017b} the notion of a uniformly rotating `pattern frame' as the rest frame of the dynamo magnetic field was introduced. The pattern frame may    also be the systemic frame of the galaxy, in which case the absolute field rotation would  be set essentially by the parameter $\epsilon$. Generally we may think of this  pattern frame of reference as the pattern speed of the gravitational  spiral arms, and then $\epsilon$ measures the  rotation of the magnetic arms relative to this reference frame.

In our first example we set $u=w=0$ and allow the similarity class $a$ to be arbitrary. Combining the  ancillary  equation  (\ref{eq:bphi1}) with equation (\ref{eq:bphi2}) yields the equation for $\tilde A_\phi$ in this case as 
\be
(1+\zeta^2)\tilde A''_\phi-2imq\zeta\tilde A'_\phi+[K^2-m^2+imq(1+imq)]\tilde A_\phi=0.\label{eq:Aphialla}
\ee

The equations for the remaining potentials follow from equations (\ref{eq:AR}) and (\ref{eq:AZ}) in the form
\be
\tilde A_r(K^2-m^2)=(1+imq)(Kv+im)\tilde A_\phi-[K-imv+\zeta(Kv+im)]\tilde A'_\phi,\label{eq:Aralla}
\ee
and 
\be
\tilde A_z(K^2-m^2)=(1+imq)(K-imv)\tilde A_\phi+[Kv+im-\zeta(K-imv)]\tilde A'_\phi.\label{eq:Azalla}
\ee
From these the magnetic field follows from equations (\ref{eq:btilde}), (\ref{eq:bbar}) and the second of (\ref{eq:tempfields}).

The equation (\ref{eq:Aphialla}) for $\tilde A_\phi$ is solved  as a function of $\zeta$ in terms of associated Legendre functions. % in the MAPLE script "spiralexactvne0alla.mw". 
   One might think that the fields should also decline with height above the disc. However at a fixed radius, increasing height implies a  decreasing vertex angle measured from the galactic axis . Consequently a magnetic field increasing with height  is actually increasing towards the minor axis of symmetry of the galaxy, which may be physically relevant.

The boundary condition at the disc must  be treated carefully in each example. Currently equation (\ref{eq:Aphialla}) is invariant under a change of sign in $\zeta$. Hence the `cuts' in the complex plane  required by the associated Legendre functions must be chosen so as to allow this continuity in the solution. Thus the `cut' should be taken away from the real axis in the range $[-1,1]$. Outside of this range the Legendre functions are evaluated as in MAPLE, being continuous onto the cuts. 

The continuity means that the disc boundary condition is ignored (i.e. the solution is neither symmetric nor anti-symmetric) across the plane). The only boundary condition that is possible that recognizes the disc is to reflect the solution for $\zeta>0$ across the plane to $\zeta<0$. This must be done while reversing the sign  of the magnetic field so that the vertical field is continuous at the plane.  This ensures an anti-symmetric tangential field across the disc (as was also required in \cite{Hen2017b}).

We show a limited sample of the behaviour of these solutions in the figures of this section. The time is set equal to $1$, but this choice is of no importance for the geometrical behaviour of the magnetic field because the field remains self-similar at all times. The time factor merely allows us to trace the growth of the field from its value at $T=0$. The Units of $T$ follow from equation (\ref{eq:T}) once a constant and a time scale are chosen.

\begin{figure}
\begin{tabular}{cc} %This will make a two-column figure
\rotatebox{0}{\scalebox{0.5} %change the angle and scale as you need
{\includegraphics{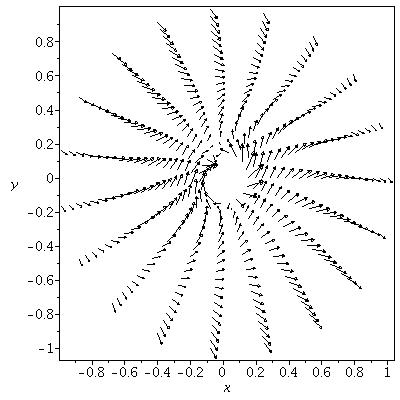}}}&
\rotatebox{0}{\scalebox{0.5} %change the angle and scale as you need
{\includegraphics{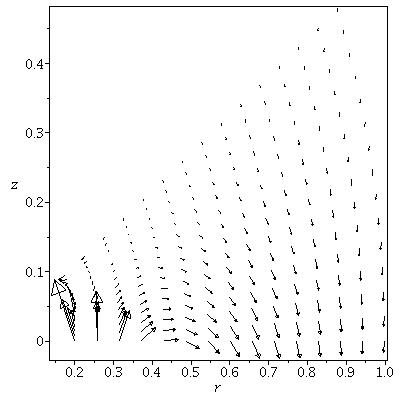}}}\\
{\rotatebox{0}{\scalebox{0.5} %change the angle and scale as you need
{\includegraphics{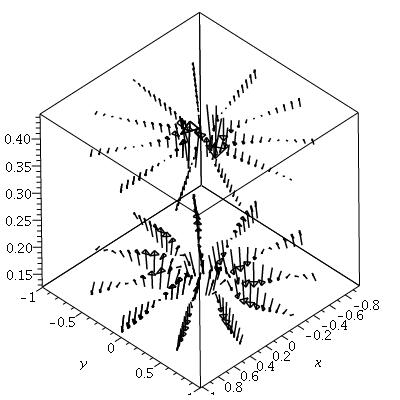}}}}&
\rotatebox{0}{\scalebox{0.5} %change the angle and scale as you need
{\includegraphics{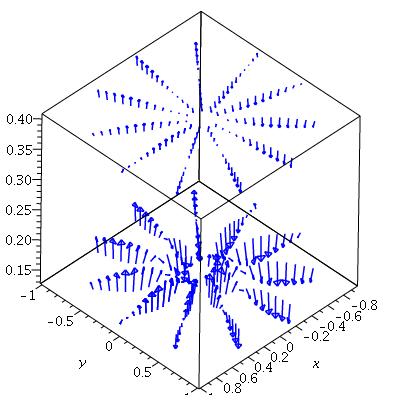}}}
\end{tabular}
\caption{ These images are for pure rotation in the pattern frame. The figure at upper left shows the magnetic field at constant $z=0.25$ (the radius of the galaxy is $r=1$), where $0.15\le r\le 1$ and $0\le \phi\le 2\pi$.  At upper right we show a poloidal cut through the same magnetic field at an angle of $\pi/4$ to the x axis where $0.2\le r\le 1$ and $0\le z\le 0.5$. The parameter vector for these figures is $\{ v,a,m,q,\epsilon,C1,C2\}$ where the  last two constants multiply each of the independent solutions for $\tilde A_\phi$. The upper two figures have the parameter vector $\{1.5,1,1,2.5,-1,0,1 \}$. 
The lower two figures show the magnetic field lines  in three dimensions with the same parameter vectors as for the upper two figures except that $a=1$ on the left and $a=2$ on the right. The absolute scaling is higher by a factor $\approx 2$ at lower left compared to lower right. }    
\label{fig:valla}
\end{figure}

\begin{figure}
\begin{tabular}{cc} %This will make a one-column figure
\rotatebox{0}{\scalebox{0.75} %change the angle and scale as you need
{\includegraphics{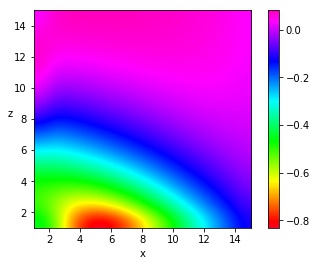}}}
\rotatebox{0}{\scalebox{0.5} %change the angle and scale as you need
{\includegraphics{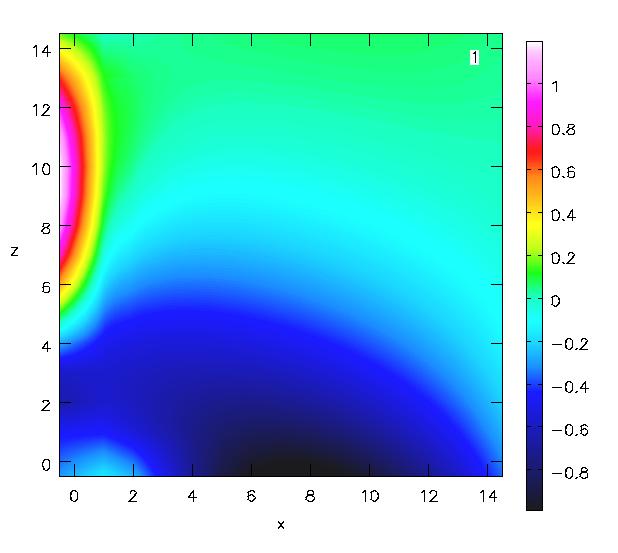}}}
%{\rotatebox{0}{\scalebox{0.5} %change the angle and scale as you need
%{\includegraphics{onehundredgravmodes.eps}}}}
\end{tabular}
\caption{ The parameter vector designates again again the quantities $\{ v,a,m,q,\epsilon,C1,C2\}$. The figure shows  the rotation measure screen in the first quadrant for the (pattern frame)  pure rotation case. On the left the parameter vector is $\{1.5,2,1,2.5,-1,1,0\}$ and the radius of the galaxy is $ 1.1$ in physical Units or $16.5$ in grid Units.  On the right we show the RM screen for the  parameter vector  $\{1.5,1,1,2.5,-1,0,1\}$, which is the vector for all of the fields in figure (\ref{fig:valla}), except the image at lower right of that figure. The signs will change in the second quadrant as they must  also change between the first and fourth and the second and third quadrants. This is true for odd modes as at present with $m=1$, but even modes will only have anti-symmetry across the plane between the first and fourth and second and third quadrants. } %Along row $2$ the RM decreases steadily through positive values by about a factor $2$.  The same is true for row $6$ except it becomes weakly negative near the edge (column 10). Near the axis well above the plane (row 9 column 2) the RM is strongly negative and decreases with radius.     
\label{fig:vallaRM}
\end{figure} 
The example at upper left shows spiral arms that are rather poorly defined, being bounded by strong fields that tend to polarization arms. The length of the vectors is calculated as $1.75$ times the ratio of the local field to grid maximum. The plot runs over $0.15\le r\le 1$  in radius and over $0..2\pi$ in azimuth.  At upper right we show a poloidal section  through the same magnetic field as at left. Unlike axially symmetric fields, the spiral field does not show `X type'  behaviour. 
More typically the field loops over the spiral arm, which behaviour was also indicated in (\cite{Hen2017b}). 

The field in three Dimensions in the lower panels of figure (\ref{fig:valla}) also indicates that the field lines loop over the spiral arms and confirms the global absence of `X type' fields. At large radius the magnetic field tends to be perpendicular to the galactic plane. The image at lower left is for the same magnetic field as on the image at upper left, while the image at lower right differs only by having $a=2$.  We see that the principal effect of the different choice of `class' (from constant velocity to constant specific angular momentum) is to cause the magnetic field to decline more rapidly with height above the plane. Moreover, the field at small radius strengthens with height when $a=1$, but is much weaker at small radius with $a=2$. Such sensitivity to a globally conserved quantity will be of interest when fitting observations.

The rotation measure screen (RM)\footnote{The RM screen is  the integral of the parallel component of the magnetic field along the line of sight, assuming a uniform density of relativistic particles.}   on the right in figure (\ref{fig:vallaRM})  shows sign changes mainly with height (resulting in `parity inversion' relative to the sign change across the disc, although there is a considerable change in RM amplitude with radius. The absolute value is subject to an arbitrary constant factor, but the relative variation is physical. The other quadrants can be generated by recalling the antisymmetry across the disc and, for odd modes, making the second and first quadrants antisymmetric. They are symmetric for the even modes. This RM screen corresponds to three of the images shown in figure (\ref{fig:valla}). There is substantial difference in detail with the RM screen on the left near the axis, which image has the same parameter set as that at lower right, except that  the class is $a=2$. As chosen both screens are negative near the disc, but become weakly positive with height. In the RM screen on the right the field is strongly positive at height at small radius, but this is no longer the disc region in most galaxies.

\subsection{Only Outflow or Accretion in the Pattern Frame}

In this section we restrict ourselves to $a=1$ and $u=v=0$ in the pattern frame.  This allows us to study outflow from or accretion onto the galactic disc, which is an important observational question. 

The combination of the auxillary equation (\ref{eq:bphi1}) with  (\ref{eq:bphi2}) now yields (the algebra can also be carried out directly from equations (\ref{eq:explicitA}) following the procedure outlined in general above) for $\tilde A_\phi$ 
\be
(1+w^2+\zeta^2)\tilde A''_\phi+2(Kw-imq\zeta)\tilde A'_\phi+[K^2-m^2(1+q^2)-im(w-q)]\tilde A_\phi=0,\label{eq:Aphiwa1}
\ee
where now 
\be
K=K(a=1;v=0)\equiv 1+im\epsilon.\label{eq:K10}
\ee
The equation is not invariant under a change in sign of $\zeta$ and $w$.  However  we will have to reflect the solution at $\zeta>0$ across the equatorial plane (with a sign change) in order to create a symmetrical relation  across the disc.  The field is therefore anti-symmetric across the disc, and even or odd in the second quadrant according as $m$ is even or odd. The solution is given in terms of hypergeometric functions. 
%in the MAPLE script "temporalcasespiralaltgenW.mw". 
We use the MAPLE default cuts in the complex plane since these are continuous onto the cut from above. There are conditions for the convergence of the hypergeometric series however, With $\epsilon<0$ these reduce to $\zeta^2<3(1+w^2)$, which normally allows the halo to be covered adequately.

\begin{figure}
\begin{tabular}{cc} %This will make a two-column figure
\rotatebox{0}{\scalebox{0.5} %change the angle and scale as you need
{\includegraphics{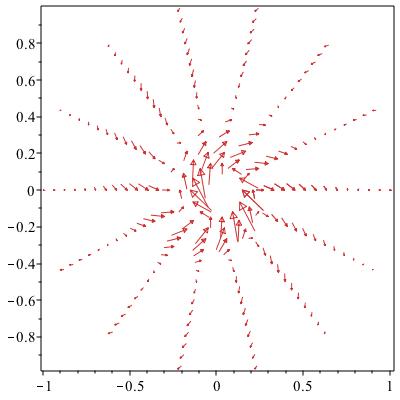}}}&
\rotatebox{0}{\scalebox{0.5} %change the angle and scale as you need
{\includegraphics{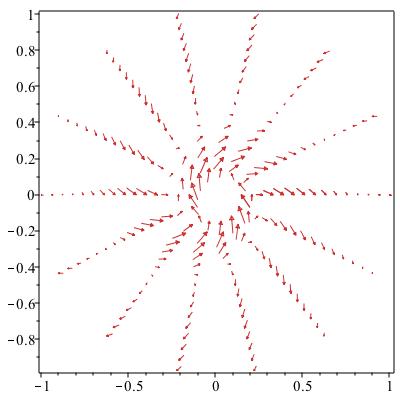}}}\\
{\rotatebox{0}{\scalebox{0.5} %change the angle and scale as you need
{\includegraphics{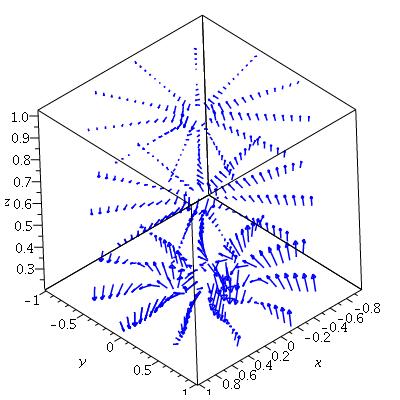}}}}&
\rotatebox{0}{\scalebox{0.75} %change the angle and scale as you need
{\includegraphics{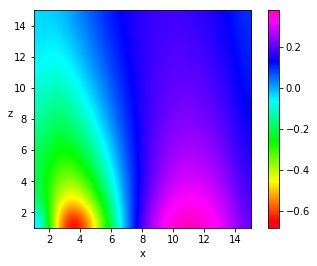}}}
\end{tabular}
\caption{These images are for the case with only $w\ne 0$ but positive and $a=1$.  At upper left the magnetic field vectors are shown on the conical surface $\zeta=0.5r$, while at upper right the field vectors are shown on a low vertical cut $z=0.15$. The radius of the galaxy is at $r=1$ in these Units. In terms of  a parameter vector $\{ m,q,\epsilon,w,T,C1,C2 \}$, these plots have the vector $\{1,2.5,-1,2.0,1,1,0\}$. The radius runs over $0.15\le r \le 1$ in each case. Vectors are a fraction of the average at each point, with the maximum vector $0.5$ times the average value. The figure at lower left shows different slices in 3D for the same parameter vector as the previous images, but over the range $0.25\le r\le 1$. At lower right we show the rotation measure screen in the first quadrant. The other quadrants may be generated  as described in figure (\ref{fig:vallaRM}). }  
\label{fig:wa1}
\end{figure}

The equations for the remaining potentials may be found from equations ((\ref{eq:AR}),(\ref{eq:AZ})) in the explicit forms

\be
[K^2-m^2(1+w^2)]\tilde A_r=[im(1+w^2)(1+imq)-K^2w]\tilde A_\phi-(1+w^2)(K+im\zeta)\tilde A'_\phi,\label{eq:Arwne0}
\ee
and
\be
[K^2-m^2(1+w^2)\tilde A_z=K[1+imq+imw]\tilde A_\phi-[K\zeta-im(1+w^2)]\tilde A'_\phi.\label{eq:Azwne0}
\ee
The dynamo magnetic field now follows from equation (\ref{eq:btilde}). We show again some examples with simple parameter choices in figure (\ref{fig:wa1}).

%The magnetic spiral arms are better defined compared to the previous example where there was only rotation in the pattern frame.  This enhancement with vertical outflow has been remarked on elsewhere (\cite{Hen2017b}),\cite{Hen2017})).  

\begin{figure}
\begin{tabular}{cc} %This will make a two-column figure
\rotatebox{0}{\scalebox{0.5} %change the angle and scale as you need
{\includegraphics{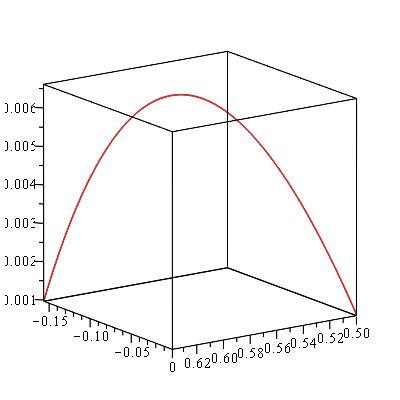}}}&
\rotatebox{0}{\scalebox{0.5} %change the angle and scale as you need
{\includegraphics{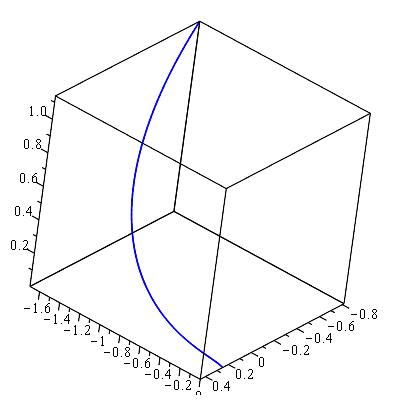}}}
%{\rotatebox{0}{\scalebox{0.4} %change the angle and scale as you need
%{\includegraphics{zerovelfieldplotm+2s1+1+.eps}}}}&
%\rotatebox{0}{\scalebox{0.4} %change the angle and scale as you need
%{\includegraphics{zerovelfieldplotm-2s1-1.eps}}}
\end{tabular}
\caption{ The closed magnetic field loop at left is for the same parameter set as in previous figures for $a=1$ and only $w\ne 0$. It begins at $\{r,\phi,z\}$ =$\{0.5,0,0.001\}$  and returns to the plane after looping in the spiral arm. The loop is very close to the plane with maximum at perhaps $60$ pc. The field line on the right is also for the same parameter set but it begins closer to the centre at $\{r,\phi,z\}$ =$\{0.25,0,0.001\}$.  We see that this line descends (the field line is pointing downwards) from great heights while crossing over the centre of the galaxy.}    
\label{fig:fieldloopsW}
\end{figure}

In figure (\ref{fig:wa1}) the spirals are much better defined than they were in figure (\ref{fig:valla}), although they are strongest at small disc radii. There is a tendency towards polarization arms at large radii leading to rather truncated spiral arms.
The  magnetic field may appear uniform near the centre of the galaxy. The spirals will continue in principle, {\it but at finite resolution this  behaviour might be seen as a `magnetic bar' from which the short magnetic spirals begin}.  Our examples  have the parameter set $\{ m,q,\epsilon,w,T,C1,C2\}=\{1,2.5,-1,2,1,1,0\} $over $0.15\le r\le 1$ and $0\le \phi \le 2\pi$. We  look at the surface of cones by setting  $z=0.5r$ yielding  $\zeta=0.5$. 
The field line structure is very markedly distributed in loops over the arms as in the previous example. This may be detected  in the cube at lower left  and is confirmed in figure (\ref{fig:fieldloopsW}).  At small radius the field lines continue to great heights without looping as is  seen on the right in figure (\ref{fig:fieldloopsW}). 

Although the projected field in an edge-on galaxy would be inclined to the disc at moderate radii, the cube at lower left  of figure (\ref{fig:wa1}) shows the field lines pointing towards the minor axis rather than away \cite{Kr2015}. As above this probably indicates the necessity  of the $m=0$ dynamo fields in order to produce `X type' dynamo magnetic fields.

The rotation measure screen is shown in the first quadrant at lower right, but the other quadrants may be generated by imposing anti-symmetry across the plane and either antisymmetry or symmetry across the minor axis depending on odd or even modes. The RM changes sign mainly in radius, which suggests recourse to an $m=0$ axially symmetric component to achieve `parity inversion' with height.

In figure (\ref{fig:fieldloopsW}) we show  on the left a magnetic field line that loops very close to the plane inside the magnetic spiral. The parameters are the same as for the other figures of this section. On the right we show a field line starting at smaller radii, but otherwise having the same set of parameters 
as elsewhere. The field line is extending to great heights and crosses over the centre of the galaxy.   

\begin{figure}
\begin{tabular}{cc} %This will make a two-column figure
\rotatebox{0}{\scalebox{0.5} %change the angle and scale as you need
{\includegraphics{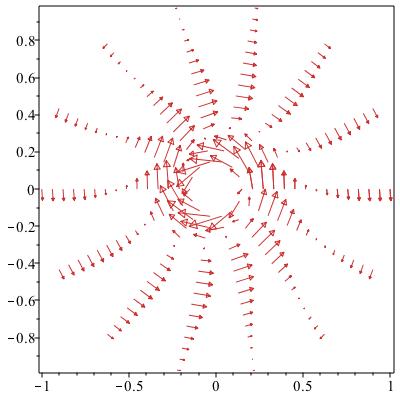}}}&
\rotatebox{0}{\scalebox{0.5} %change the angle and scale as you need
{\includegraphics{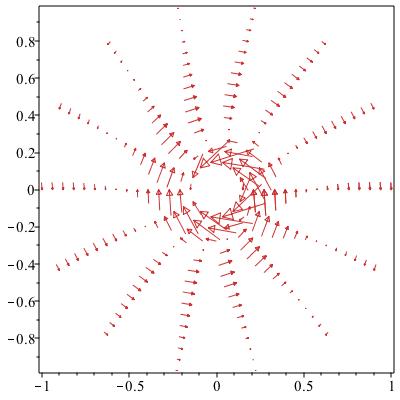}}}\\
{\rotatebox{0}{\scalebox{0.5} %change the angle and scale as you need
{\includegraphics{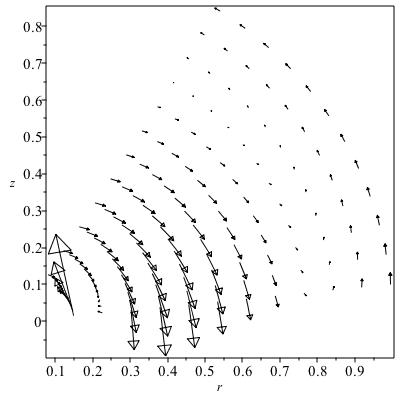}}}}&
\rotatebox{0}{\scalebox{0.7} %change the angle and scale as you need
{\includegraphics{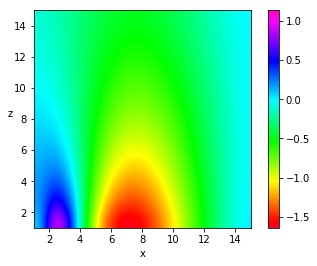}}}
\end{tabular}
\caption{The image at upper left of the figure shows a cut through the halo at $z=0.15$. The vertical velocity is $-2$ so that there is accretion onto the disc. The other parameters are the same as in figure (\ref{fig:wa1}), including the range of radius and $a=1$. At upper right we show the spiral structure on the cone $\zeta=0.25r$ over the same range in radius. Once again the only change is that the vertical flow is now inflow with $w=-2$.  At lower left we show a poloidal section at $\phi=\pi/4$. At lower right we show the RM screen for accretion  ($w=-2$) with the same parameters otherwise. }    
\label{fig:accretionW}
\end{figure}

\newpage
The magnetic field is in fact stronger and the spirals are better defined under accretion ($w<0$) \cite{Hen2017}. This is demonstrated in figure (\ref{fig:accretionW}).

Figure (\ref{fig:accretionW}) shows a dramatic improvement of the magnetic spiral structure relative to the outflow results of figure (\ref{fig:wa1}), both at a constant cut in $z$ and on the face of a cone.  At lower left we show a poloidal section at $\phi=\pi/4$ for the same accretion parameters. The field again loops above the disc, crossing over the centre of the galaxy (we have checked that the field at $\phi=5\pi/4$ has the opposite sign). The projected magnetic field is not  `X-shaped'. We have not corrected for  the  internal Faraday rotation of the locally produced emission in the presumed projections. 

The RM screen for the same accretion case is imaged at lower right of the figure. Although the amplitudes vary considerably, most of the high halo is of uniform sign. Only near the plane and near the minor axis is there a strong sign change. Rapid variation in the magnetic field is also detectable in the poloidal section at lower left of the figure. A detailed RM model would require assuming the distribution of the relativistic electrons and ideally, performing RM synthesis (or the equivalent). We are only  calculating an RM screen, due solely to the magnetic field structure while assuming a constant electron density. Should both of these increase strongly with decreasing radius, our calculation mainly reflects conditions near the tangent point of the line of sight to a given circle in the disc.

\begin{figure}
\begin{tabular}{cc} %This will make a two-column figure
\rotatebox{0}{\scalebox{0.75} %change the angle and scale as you need
{\includegraphics{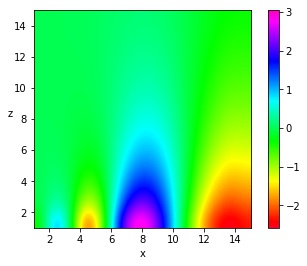}}}&
\rotatebox{0}{\scalebox{0.5} %change the angle and scale as you need
{\includegraphics{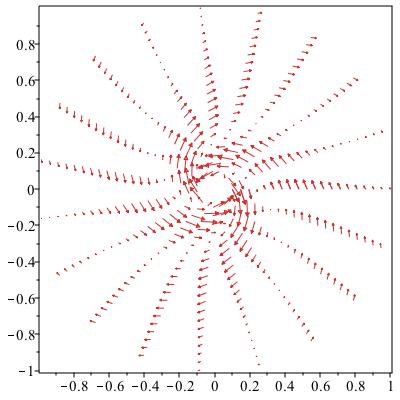}}}
%{\rotatebox{0}{\scalebox{0.4} %change the angle and scale as you need
%{\includegraphics{zerovelfieldplotm+2s1+1+.eps}}}}&
%\rotatebox{0}{\scalebox{0.4} %change the angle and scale as you need
%{\includegraphics{zerovelfieldplotm-2s1-1.eps}}}
\end{tabular}
\caption{ The figure shows the RM screen in the first quadrant for the parameter set $\{ m,q,\epsilon,w,T,C1,C2 \}$=$\{2,2.5,-1,-2,1,1,0\}$ in the left panel. The sign change is now more frequent. The right panel is a cut at $z=0.15$ over the radial range $\{0.1,1\}$ for the same parameters, except that $q=1$.  }    
\label{fig:accretionm2RM}
\end{figure}

Finally for this case we show  in figure (\ref{fig:accretionm2RM}) on the left panel the RM screen for the same accretion case as in figure (\ref{fig:accretionW}), except that $m=2$. 

The RM  screen is  more structured because of the increased number of magnetic spiral arms. They continue from the disc into the halo although much of the activity is at small $\zeta$ (but moderate height). Despite this being rather an arbitrary example, there is some resemblance to the discovery announced in \cite{SPK2016} for the CHANG-ES galaxy NGC 4631. This type of oscillation in the RM was predicted  in \cite{Hen2017b} for modal solutions, and is confirmed here. The lack of resistivity has not changed this behaviour very much.

On the right hand panel of the figure we show a cut of the same example with accretion, but with a $45^\circ$ pitch angle.  This may be compared to the upper right panel in figure (\ref{fig:accretionW})with pitch angle $68.2^\circ$.  Similar behaviour is shown in the lower right panel of figure 1 in \cite{Hen2017b}, but again for  pitch angle $68.2^\circ$. Although we have made no attempt at a proper fit, these figures show a resemblance to the observations of NGC4736 reported in figure 2 of \cite{CB2008}. The current example is for the class $a=1$ with infinite conductivity, while the example in \cite{Hen2017b} contains finite resistive diffusion and is for the similarity class $a=2$.  The velocity field, helicity and diffusion(in (\cite{Hen2017b})) all have global variations consistent with the specified $a$.  This particular galaxy is unique only in that it shows a two-armed mode extending well into the galactic centre independent of gravitational spirals. Many similar cases of magnetic spiral arms exist \cite{Beck2015}, \cite{WI2015}.

It is therefore clear that there exist  spiral modes of the dynamo equations that can fill a galactic disc, as is often observed \cite{Beck2015}.  This is possible without the direct  influence of the normal gravitational arms, although these can be fitted into the scale invariant picture to perhaps create the scale invariant velocity field (see appendix in \cite{Hen2017b}). But this is not required for NGC4736. These spiral modes extend into the halo where they will produce oscillations in the RM on each side of the minor axis. Strong `X-type' magnetic fields may require the presence of the axially symmetric mode.

Our analysis can not explain the origin of the pitch angle, which remains a parameter consistent with the scale invariance. Fortunately the general behaviour is not greatly sensitive to this value.
The spiral structure of these examples continues at all scales with varying amplitude, by construction. However it is likely that magnetic rings  or bars would be observed near the centre of these galaxies given limited resolution. A ring structure is particularly noticeable at upper left in figure (\ref{fig:accretionW}) and indeed also at lower right in figure 1 of \cite{Hen2017b}.
We intend to apply our solutions to such cases in detail in a subsequent study, in order to extract the most likely set of parameters.

\section{Conclusions}
We have only begun with this article to analyze the different possibilities for scale-invariant spiral magnetic fields. We intend to exhibit the type of calculations we have used on line\footnote{A web site is being prepared, whose location may be found by asking at henriksn@astro.queensu.ca}, so that anyone interested may pursue the investigations and comparisons with data. Among other restrictions, we have only looked at the $m=1$ spiral mode.  It is nevertheless clear that:

1. Disc spiral magnetic fields are `lifted' into the halo on cones, assuming scale invariance.

2. The edge-on projection of these same magnetic spirals typically do not  give `X type' magnetic field lines. An axially symmetric mode may be necessary to establish this behaviour.

3. Anti-symmetry of the tangential magnetic field is likely on crossing the disc. Moreover there is a modal dependence of the symmetry that holds on crossing the minor axis of the galaxy.  Even modes have even symmetry across this axis, while odd modes have odd symmetry. One must  remember  that such a spiral magnetic field may coexist with an axially symmetric magnetic field component. The axially symmetric modes may be either symmetric or anti-symmetric  on crossing the galactic disc with respect to the (line of sight) azimuthal field \cite{Hen2017}.

4.  The spiral structure can be quite complex in its variation with radius. At smaller radii there is a tendency for the magnetic spiral to be seen as an apparent `ring'  or `bar'.  Some magnetic spiral structures are `polarization arms' wherein the field is not necessarily directed along the spiral. Near the plane the magnetic field frequently {\it loops over the magnetic arms}. On larger scales it can loop over the centre of the galaxy. This is a product of the dynamo and is not related to the Parker instability.

5. The rotation measure screen (RM)  produced by these spiral modes may show sign changes  with height on the same side of the minor axis (`parity inversion'), but more typically it does not. This effect seems also to be primarily a property of axially symmetric modes \cite{Hen2017}. The symmetry exhibited on crossing the disc will depend on the azimuthal magnetic field symmetry and hence on the axially symmetric component combined with the spiral mode structure.
Such symmetry is being detected in the CHANG-ES data (e.g. \cite{CMP2016}), and was already discussed in \cite{Hen2017b}.

6. The self-similar time dependence is fixed by the presence of a global constant whose Dimensions determine the scale invariant class $a$. This has not been studied extensively here, but it might be integrated into the evolution and conserved properties of the galaxy together with the corresponding axially symmetric model \cite{HWI2018}. The time dependence also  allows for a relative rotation between the gravitational  spiral arms and the magnetic  spiral arms.

It is left for future work to combine this analysis with a scale invariant treatment of the halo magnetohydrodynamics. However it is not impossible that the  scale invariant form of the velocity used presently is relevant physically. We know that complex systems tend to become scale free asymptotically, and this  symmetry can be applied to the familiar gravitational arms \cite{Hen2017b}. Moreover assuming scale invariant symmetry for the mean field is formally  consistent with a scale invariant treatment of the sub scale turbulence (e.g. \cite{Hen2015}).

\section{Acknowledgements}

\newpage
%\begin{figure}%[p]
%\begin{tabular}{cc} %This will make a two-column figure
%\rotatebox{0}{\scalebox{0.65} %change the angle and scale as you need
%{\includegraphics{zerovelfield3d1.jpg}}}&
%\rotatebox{0}{\scalebox{0.65} %change the angle and scale as you need
%{\includegraphics{zerovelfield3d2.jpg}}}\\
%{\rotatebox{0}{\scalebox{0.4} %change the angle and scale as you need
%{\includegraphics{zerovelfieldplotm+2s1+1+.eps}}}}&
%\rotatebox{0}{\scalebox{0.4} %change the angle and scale as you need
%{\includegraphics{zerovelfieldplotm-2s1-1.eps}}}
%\end{tabular}
%\caption{}    
%\label{fig:vzeroField3d}
%\end{figure}

%\begin{figure}%[p]
%\begin{tabular}{cc} %This will make a one-column figure
%\rotatebox{0}{\scalebox{0.6} %change the angle and scale as you need
%{\includegraphics{Fvcombinedspacecurve.jpg}}}
%\rotatebox{0}{\scalebox{0.6} %change the angle and scale as you need
%{\includegraphics{zerovelRMm22nd.jpg}}}
%{\rotatebox{0}{\scalebox{0.5} %change the angle and scale as you need
%{\includegraphics{onehundredgravmodes.eps}}}}
%\end{tabular}
%\caption{ The solid curve shows a loop above the galactic disc calculated for the $m=-1$, $s1=-1$ mode. The dashed curve is for the $m=+1$, $s1=-1$ mode. Both curves are started at $r=0.75$, $\phi=3\pi/4$ and $z=0.0015$. The value  $q=0.4$ for both loops. }    
%\label{fig:vonlyloops}
%\end{figure} 

\label{lastpage}
\end{document}